\documentclass[12pt,preprint]{aastex}

\newcommand{\eg}{e.g., }

\newcommand{\FxFopt}{F_{\rm X}/F_{\rm opt}}

\begin{document}

\shorttitle{Extreme X-Ray / Optical Ratio Sources}

\title{A Possible New Population of Sources with Extreme X-Ray / Optical
Ratios
$^{1,2}$}

\footnotetext[1]{$^{,2}$Based on observations obtained with the NASA/ESA {\it Hubble
Space Telescope}, which is operated by the Association of Universities for
Research in Astronomy, Inc., under NASA contract \hbox{NAS 5-26555}, and
at the European Southern Observatory, Chile
(164.O-0561, 169.A-0725, 267.A-5729, 66.A-0451, 68.A-0375).}

\author{Anton M. Koekemoer\altaffilmark{3},
D. M. Alexander\altaffilmark{4,5},
F. E. Bauer\altaffilmark{4},
J. Bergeron\altaffilmark{6},
W. N. Brandt\altaffilmark{4},
E. Chatzichristou\altaffilmark{7},
S. Cristiani\altaffilmark{8},
S. M. Fall\altaffilmark{3},
N. A. Grogin\altaffilmark{9},
M. Livio\altaffilmark{3},
V. Mainieri\altaffilmark{11},
L. Moustakas\altaffilmark{3},
P. Padovani\altaffilmark{3,10},
P. Rosati\altaffilmark{11},
E. J. Schreier\altaffilmark{12},
C. M. Urry\altaffilmark{7}
}

\altaffiltext{3}{
	Space Telescope Science Institute,
	3700 San Martin Drive, Baltimore, MD 21218, USA}

\altaffiltext{4}{
	Penn State University,
	525 Davey Lab, University Park, PA 16802 USA}

\altaffiltext{5}{
	Institute of Astronomy,
	Madingley Road, Cambridge, CB3 0HA, UK}

\altaffiltext{6}{
	Institut d'Astrophysique de Paris,
	98bis Bd Arago, F-75014 Paris, France}

\altaffiltext{7}{
	Department of Physics (JWG 460), Yale University,
	P.O. Box 208121, New Haven CT 06520-8121, USA}

\altaffiltext{8}{
	INAF-Osservatorio Astronomico,
	via Tiepolo 11, I-34131 Trieste, Italy}

\altaffiltext{9}{
	Department of Physics and Astronomy,
	Johns Hopkins University,
	Baltimore, MD 21218, USA}

\altaffiltext{10}{
	ESA Space Telescope Division}

\altaffiltext{11}{
	European Southern Observatory,
	Karl-Schwarzschild-Str. 2, Garching D-85748, Germany}

\altaffiltext{12}{
	Associated Universities, Inc.,
	1400-16th St, NW, Ste 730, Washington, DC 20036, USA}

%
%
%
%
%
%
%
%
%
%
%
%
%

%
\slugcomment{ApJ Letters, Accepted 2003 May 29}

\begin{abstract}

We describe a possible new class of X-ray sources that have robust detections
in ultra-deep {\it Chandra} data, yet have no detections at all in our deep
multi-band GOODS {\it Hubble Space Telescope} ({\it HST}) ACS images, which
represent the highest quality optical imaging obtained to date on these fields.
These extreme X-ray~/ Optical ratio sources ({\it ``EXO''}s) have values of
$\FxFopt$ at least an order of magnitude above those generally found for other
AGN, even those that are harbored by reddened hosts. We thus infer two possible
scenarios: (1)~if these sources lie at redshifts $z \lesssim 6$, then their
hosts need to be exceedingly underluminous, or more reddened, compared with
other known sources; (2)~if these sources lie above $z \sim 6-7$, such that
even their Lyman-$\alpha$ emission is redshifted out of the bandpass of our ACS
$z_{850}$ filter, then their optical and X-ray fluxes can be accounted for in
terms of relatively normal $\sim L_*$ hosts and moderate-luminosity AGN.

\end{abstract}



\section{Introduction}

A key question in astrophysics concerns the evolution of active galactic nuclei
(AGN) during the ``quasar epoch'' ($z \sim 2 - 3$) and at higher redshifts,
where their space density declines
	(Fan et al. 2001, 2003;
	Barger et al. 2003).
Their evolution appears to track the star formation rate
	(\eg Steidel et al. 1999),
thereby suggesting an empirical link between galaxy growth and AGN fuelling.
A connection between galaxies and AGN is also suggested by the
black hole~/ bulge mass relationship
	(Ferrarese \& Merritt 2000;
	Gebhardt et al. 2000).

A powerful tool to investigate the physical nature of such relationships is AGN
X-ray emission, which above a few keV can penetrate obscuration around AGN.
Ultra-deep X-ray surveys with {\it Chandra} on the Hubble Deep Field North
(HDF-N;
	Brandt et al. 2001),
and Chandra Deep Field South
	(CDF-S; Giacconi et al. 2002)
are sufficiently sensitive to reveal AGN beyond the tentative reionization epoch
($z \sim 6-7$,
	Fan et al. 2001, 2003),
where optical information is no longer available. These surveys also probe more
numerous, lower-luminosity AGN at $z \sim 3 - 6$, where ultra-deep optical and
near-IR imaging can constrain their properties.

In this paper we describe a sample of sources with extreme X-ray / Optical flux
ratios (or {\it ``EXO''}s), that are robustly detected ($25 - 89$ counts) in
the 2~Msec HDF-N and reprocessed 1~Msec CDF-S main catalogs
	(Alexander et al. 2003).
They are also detected in near-IR $JHK$ imaging, but completely undetected in
our deep GOODS {\it HST/}ACS survey, which to date is the most sensitive and
detailed optical imaging of these fields
	(Giavalisco et al. 2003).
We discuss various possible interpretations for these objects. Throughout this
paper we adopt $H_0 = 70$~km~s$^{-1}$~Mpc$^{-1}$, $\Omega_{\rm M} = 0.3$, and
$\Omega_\Lambda = 0.7$.

\section{Observations and Sample Description}

We began by matching the X-ray catalogs
	(Alexander et al. 2003)
to our ACS $z_{850}$ catalogs
	(Giavalisco et al. 2003);
details are in
	Koekemoer et al. (2003, in preparation)
and
	Bauer et al. (2003, in preparation).
The matched sources are mostly moderate-luminosity AGN at $z \sim 0.5 - 4$,
or star-forming galaxies at $z \lesssim 0.5 - 1$
	(\eg Hornschemeier et al. 2001;
	Schreier et al. 2001;
	Alexander et al. 2001;
	Koekemoer et al. 2002).

The remaining sources were inspected in detail by overlaying X-ray contours on
the $z_{850}$ and the combined $B_{435}$+$V_{606}$+$i_{775}$+$z_{850}$ images.
Most of these had faint optical counterparts, below the formal 10-$\sigma$
$z_{850}$ catalog threshold or undetected in $z_{850}$ but detected in another
band (perhaps high equivalent width Lyman-$\alpha$ emitters). Detection in
$z_{850}$ or bluer bands suggests these sources are at $z \lesssim 6$
	(Barger et al. 2003;
	Cristiani et al. 2003).

Finally, there remained sources with no ACS counterparts within $2\farcs4$
($\gtrsim 10$ times the positional uncertainty) in $z_{850}$ and the combined
$B_{435}$+$V_{606}$+$i_{775}$+$z_{850}$ images, despite robust detections in
the X-ray main catalogs ($25 - 89$ counts). We focus on CDF-S where we have
complete ACS and near-IR coverage; HDF-N sources will be discussed separately
	(Koekemoer et al. 2003, in preparation).
Five of these 7 CDF-S sources have fairly well behaved X-ray exposure maps; the
other two are still detected in $JHK$ thus are likely also real. The sources are
also at a wide range of off-axis angles in the {\it Chandra} image. We computed
$z_{850}$ upper limits from the pixel r.m.s. ($0.0017 - 0.0024$~counts~s$^{-1}$~pixel$^{-1}$),
integrated over a $0\farcs2$$\times$$0\farcs2$ aperture (4$\times$4 ACS-pixel
box), to yield 3-$\sigma$ upper limits%
\footnote{Using the same zeropoints as
	Giavalisco et al. (2003).}
of $z_{850}\sim 27.9 - 28.4$ (AB magnitudes).

\section{``EXO''s: Extreme X-ray / Optical Ratio Sources}

The 7 sources with no ACS counterparts are shown in Figure~\ref{fig1}. We
carried out photometry at the X-ray positions, and present the ACS upper limits
and near-IR detections in Table~\ref{tab1}. To further investigate the nature
of these sources, we examine the relationship between their X-ray and optical
fluxes
	(\eg Maccacaro et al. 1988;
	Stocke et al. 1991).
In Figure~\ref{fig2} we plot $F_{\rm 0.5-8\,keV}$ versus $z_{850}$ for all the X-ray
sources, including the ACS non-detections, also showing lines of constant $\FxFopt$
(derived from $z_{850}$). At soft X-ray energies this ratio is complicated by
absorption and thermal gas emission
	(\eg Beuermann et al. 1999)
but these no longer dominate above a few keV, which is redshifted into the soft
band for $z \gtrsim 1-2$. Sources with $\FxFopt \gtrsim 0.1-10$ are generally
identified with AGN; these deep X-ray data also reveal normal galaxies and
starbursts
	(\eg Hornschemeier et al. 2003)
with $\FxFopt$ extending a few orders of magnitude below AGN, but at
$z_{850} \gtrsim 24$ these are lost from the X-ray sample and the AGN dominate.

Figure~\ref{fig2} also reveals that our $z_{850}$-undetected objects occupy the high
end of the $\FxFopt$ plane, above most of the detected optically faint sources. This
is interesting since it is not expected {\it a priori} that optically undetected
objects should have strong X-ray flux. We also plot our measured ACS magnitudes for
CDF-S sources reported as undetected in ground-based data by
	Yan et al. (2003),
and show Extremely Red Objects (EROs) from previous studies
	(Alexander et al. 2002;
	Stevens et al. 2003).

In Figure~\ref{fig3} we plot $\FxFopt$ against $z_{850} - K$; the normal galaxies
and starbursts with $\FxFopt \lesssim 0.1-1$ are blue, while the only objects with
$z_{850}-K > 4$ are those with high $\FxFopt$ (see also
	Brusa et al. 2002;
	Cagnoni et al. 2002).
Not {\it all} the high $\FxFopt$ objects are red; some are blue, with
$z_{850}-K \sim 0-2$, and are likely unobscured AGN. However, at high $\FxFopt$
we also find the reddest objects, with most of our undetected $z_{850}$ sources
at the extreme end with $z_{850}-K \gtrsim 4.2 - 6.2$.

\section{Possible Constituents of the ``EXO'' Population}

We now examine various possibilities for these sources in order to explain:
(1)~deep optical non-detection; (2)~robust X-ray detection; and (3)~extremely
red colors. We rule out main-sequence stars as these typically have
$\FxFopt \sim 10^{-5} - 0.1$
	(e.g., Stocke et al. 1991).
More exotic galactic sources include low-mass X-ray binaries, cataclysmic
variables, or neutron stars, which generally have
$L_X \sim 10^{30} - 10^{35}$~erg~s$^{-1}$
	(e.g., Verbunt \& Johnston 1996 and references therein).
At distances $\sim 10 - 100$~kpc they would have X-ray fluxes
$\lesssim 10^{-14}$~erg~s$^{-1}$~cm$^{-2}$, consistent with what is observed.
However, the number of sources in our small survey area would imply total
counts $\sim 10^4 - 10^5$ times above what is known for our galaxy
	(e.g. Howell \& Szkody 1990;
	Alexander et al. 2001).
Thus we turn to an extragalactic origin for the {\it EXO}s.

Low-luminosity galaxies can host relatively luminous X-ray sources, for example
the local dwarf NGC~4395 ($M_B \sim -16.5$) with a central X-ray source
$L_{\rm 2-10 keV} \sim 3 \times 10^{38}$~erg~s$^{-1}$
	(Ho et al. 2001).
However, such objects do not have the extremely red colors of our sources.
Moreover, our $z_{850}$ limits would imply a distance modulus $\gtrsim 45.3$~mag,
hence $L_X \gtrsim 3 \times 10^{42} - 8 \times 10^{43}$~erg~s$^{-1}$, at least
$10^4$ times above those for such dwarfs locally. It would be highly interesting
if the {\it EXO}s were low-luminosity dwarfs with such extreme AGN, but since
we know of no local analogs, we explore other alternatives.

\subsection{Evolved Galaxies}

A possible alternative is that the {\it EXO}s may be at moderate redshifts and
extremely evolved. At $z \sim 1,2,3,4,5,6$, the oldest possible populations
(5.7, 3.2, 2.1, 1.5, 1.2, 0.9~Gyr) would have observed colors
$z_{850}-K \sim 1, 4, 4.5, 3.5, 3.2, 3$ respectively, for a single-burst
model with 0.3 solar metallicity
	(Charlot \& Bruzual 2003, in preparation).
At least $\sim$$1 - 2$ magnitudes of extinction would be required since we
observe no UV excess. These properties are comparable to those derived for
other X-ray ERO's (e.g.,
	Alexander et al. 2002;
	Brusa et al. 2002;
	Stevens et al. 2003).
However, it is interesting to note that the $K$-band magnitudes are $\sim 100$
times too low compared to what is needed to place these objects on the
$M_{BH}-\sigma$ relation (see
	Woo \& Urry 2002).

\subsection{Dusty Galaxies}

We next explore the amount of dust needed to achieve the high $z_{850}-K$
colors. If these sources were local ($z \lesssim 0.1$), then
$z_{850}-K \gtrsim 4.5$ would imply $A_V \gtrsim 10 - 14$
	(Rieke \& Lubofsky 1985;
	Calzetti 1997),
depending on the stellar populations. This would require a column
$N_H \gtrsim 2 - 5 \times 10^{22}$~cm$^{-2}$ 
ratio, which is too high for the observed soft X-ray fluxes unless we invoke a
highly interesting scenario where the AGN is unobscured while the rest of the
galaxy is completely obscured.

At higher redshifts, the $A_V$ limits decrease since $z_{850}$ and $K$ sample
bluer rest-frame wavelengths. Between $z \sim 1 - 2.5$ we infer
$A_V \gtrsim 5 - 2$ respectively, comparable to similar objects discussed by
	Alexander et al. (2002);
	Smail et al. (2002);
	Stevens et al. (2003), and
	Yan et al. (2003).
Again it is interesting that the {\it EXO} host luminosities at these redshifts,
from their $K$ magnitudes, would place them at least 2 orders of magnitude away
from the $M_{BH}-\sigma$ relation implied by their X-ray fluxes, suggesting
atypical accretion rates
	(Woo \& Urry 2002).

\subsection{High-Redshift Lyman Break Galaxies}

Finally we investigate the possibility that these sources are at high enough
redshifts to move the Lyman break out of the $z_{850}$ filter.
	Yan et al. (2003)
investigated similar possibilities for their CDF-S sources undetected in
ground-based $R$ imaging but detected with ISAAC. We have in fact detected their
sources in our ACS $z_{850}$ data (Figure~\ref{fig2}) and agree with them that
those sources are unlikely to be at $z \gtrsim 5$. However, their sources are
the only ones with $\FxFopt$ comparable to our {\it EXO}s. Thus, if the
{\it EXO}s belong to the same population as the
	Yan et al. (2003)
sources, then their distance modulus is $\gtrsim 1.5 - 2$ magnitudes higher. If
the
	Yan et al. (2003)
sources are luminous E/S0 galaxies with $z \sim 1 - 2.5$, this would translate
to $z\sim 2.5 - 6$ for the {\it EXO} population. The other possibility
considered by
	Yan et al. (2003)
is that their sources may be higher redshift AGN up to $z \sim 3-5$, depending
on extinction details. If {\it EXO}'s are the high redshift extension of this
population, this would place them at $z \sim 6-10$. In this case, their host
galaxy absolute magnitudes (from our $K$-band data) would be $-$20 to $-$21,
comparable to $\sim L_*$ luminosities, while their X-ray luminosities would be
$\sim 2 - 6 \times 10^{44}$~erg~s$^{-1}$. These are not unusually high
luminosities for AGN, and may be consistent with a high-redshift extension of
moderate-luminosity AGN at $z\sim 3-5$
	(eg Cristiani et al. 2003).

\section{Conclusions}

We have examined various possible explanations for a population of obejcts that
have extreme X-ray / Optical ratios ({\it ``EXO''}s), being robustly detected
in the Chandra data in CDF-S and HDF-N, yet completely undetected in our deep
multi-band $B_{435}$+$V_{606}$+$i_{775}$+$z_{850}$ ACS images, which represent
the most sensitive optical imaging obtained to date on these fields. In
particular, the lack of detections even in the $z_{850}$ band suggests one of
two possible scenarios: (1)~if these sources lie at $z \lesssim 6-7$, then
their hosts are unusually underluminous, or more reddened, compared with
other known AGN hosts (even other ERO's); (2)~if these sources lie above
$z \sim 6-7$, such that even their Lyman-$\alpha$ emission is redshifted out of
the ACS $z_{850}$ bandpass, then their optical and X-ray fluxes can be accounted
for in terms of $\sim L_*$ hosts and moderate-luminosity AGN. Deep near-IR
spectroscopic observations will help further elucidate their nature.

\acknowledgements

Partial support for this work was provided by NASA through grant number
\mbox{HST-GO-09583.01-96A} from the Space Telescope Science Institute, which is
operated by the Association of Universities for Research in Astronomy, under
NASA contract \mbox{NAS 5-26555}. Support for this work, part of the
{\it Space Infrared Telescope Facility (SIRTF)} Legacy Science program,
was also provided by NASA through Contract Number 1224666 issued by the Jet
Propulsion Laboratory, California Institute of Technology under NASA contract
1407. We thank the referee for a speedy review and for valuable suggestions that
helped to improve this paper. AMK wishes to thank Mark Dickinson for insightful
discussions, and Harry Ferguson for thorough comments on the manuscript.
DMA acknowledges the generous support provided by a Royal Society University
Research Fellowship. DMA, FEB and WNB also wish to thank NSF CAREER award
\hbox{AST-9983783}.

\clearpage

\begin{figure*}
\epsscale{0.9}
\caption{\label{fig1}
The 7 CDF-S X-ray sources with no ACS counterparts,
also showing the SOFI $JHK$ detections.
Each panel is $15''$ on a side. Contours show $0.5-8$~keV {\it Chandra} data,
starting at 1,2,3-$\sigma$, and doubling thereafter.
For the full image, see http://www.stsci.edu/$\sim$koekemoe/goods-exo/}
\end{figure*}

\clearpage

\begin{figure}
\plotone{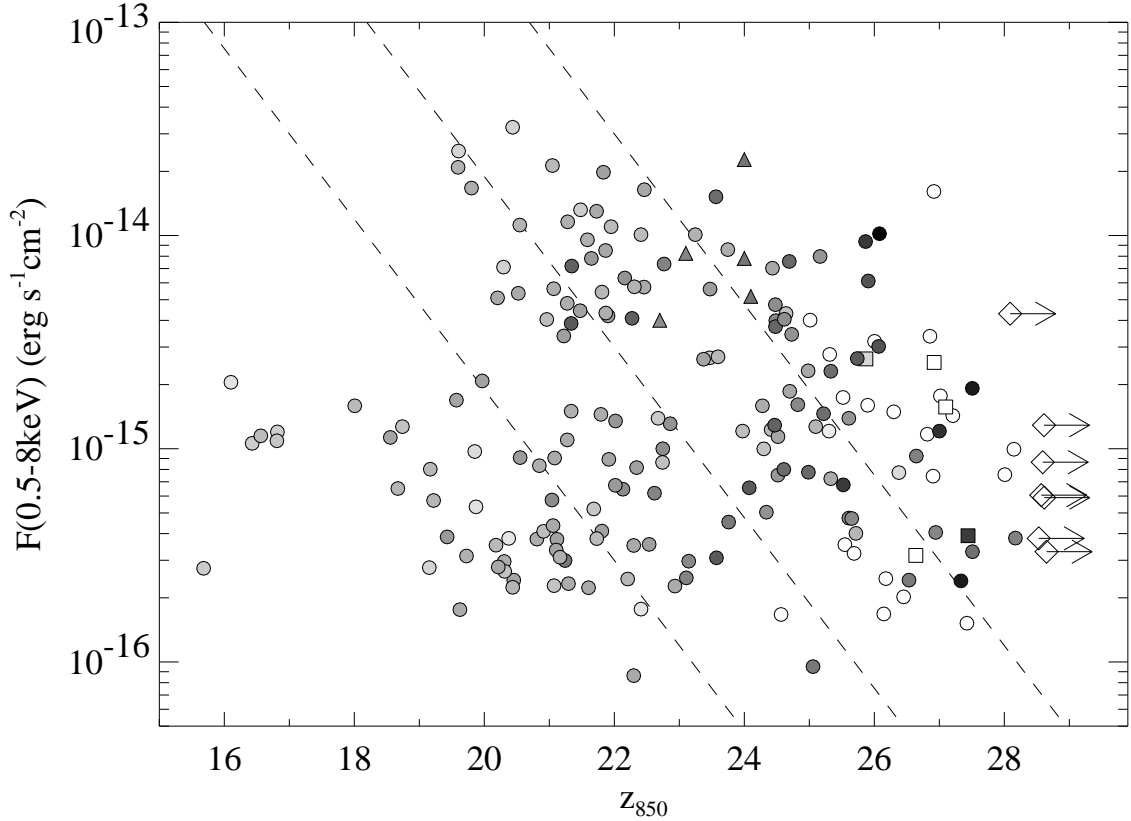}
\caption{\label{fig2}
Total X-ray flux ($0.5-8\,$keV) against $z_{850}$ magnitude for all the X-ray
sources, including those unidentified in $z_{850}$ (indicated by diamonds
with arrows). Lines indicate $\FxFopt=0.1, 1, 10$. Symbol shading shows
the $z_{850}-K$ colour of each source, ranging from light to dark for
$z_{850}-K \sim 0 - 5$. White symbols (no shading) were undetected in $K$.
Squares show our measured ACS magnitudes for sources reported as undetected
in ground-based data by
	Yan et al. (2003),
and triangles indicate X-ray selected EROs in the Lockman Hole
	(Stevens et al. 2003).}
\end{figure}

\begin{figure}
\plotone{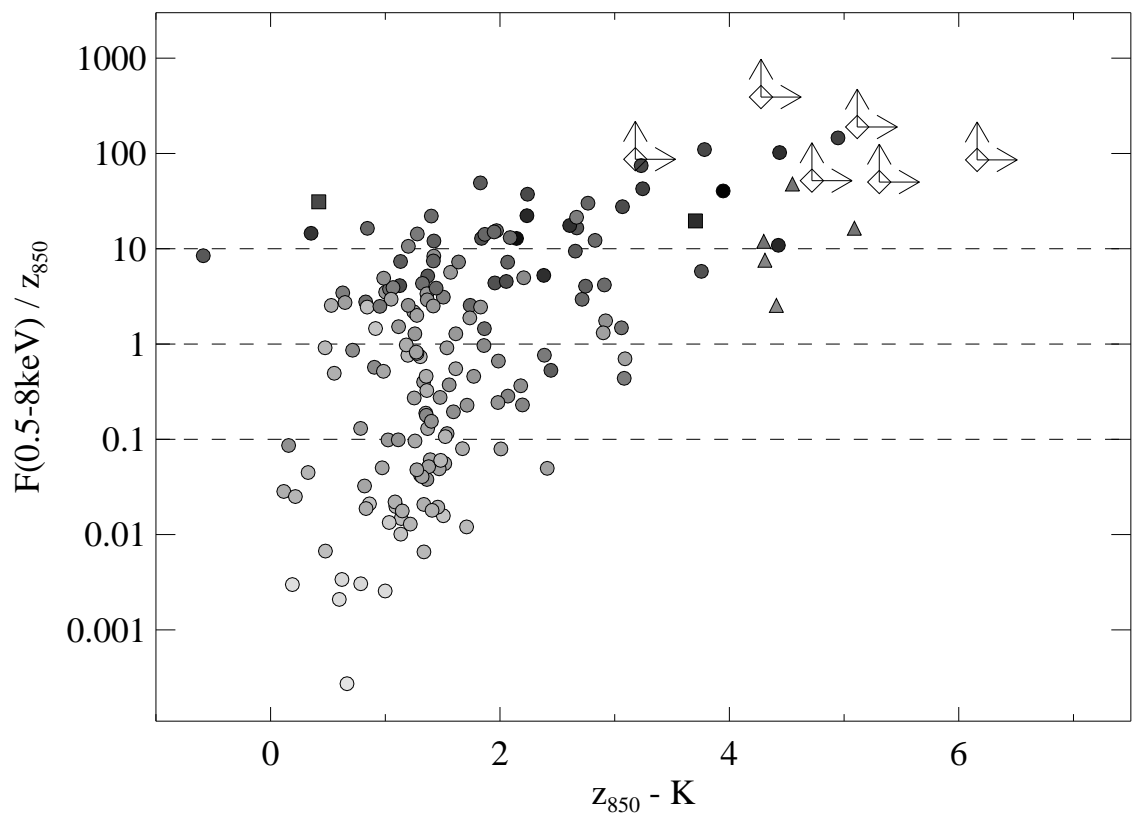}
\caption{\label{fig3}
Ratio of $F_{\rm 0.5-8\,keV}$ to $z_{850}$ for all the X-ray sources, plotted
against $z_{850}-K$. Symbol shapes are as in Figure~\ref{fig2}; however shading
represents $z_{850}$, fainter sources being darker. Normal galaxies and
starbursts ($\FxFopt \lesssim 0.1-1$) are relatively blue, and that the only
redder objects are those with higher $\FxFopt$.}
\end{figure}

\begin{deluxetable}{ccccccccccc}
\tablewidth{0pt}
\tablecaption{\label{tab1}%
	Optically Undetected X-Ray Sources}
\tablehead{%
R.A.\tablenotemark{a}
	& Dec.\tablenotemark{a}
		 	& Total counts
				& F$_{0.5-8 {\rm keV}}$
							& $z_{850}$
								& $J$
										& $H$
												& $K$		\\
(J2000)	& (J2000)	& ($0.5 - 8$ keV)	& (erg$\,$s$^{-1}$cm$^{-2}$)			}
\startdata
03 32 08.39 & $-$27 40 47.0  & 89$^{+15}_{-14}$	& $4.3\times 10^{-15}$ 	& $>$27.9& 25.2$\pm1.3$	& 23.3$\pm0.2$	& 24.8$\pm1.1$	\\
\phn03 32 08.89\tablenotemark{b}
	    & $-$27 44 24.3  & 27$^{+9}_{-8}$	& $3.8\times 10^{-16}$	& $>$28.3& $>$25.9	& 24.8$\pm0.5$	& 23.8$\pm0.6$	\\
03 32 13.92 & $-$27 50 00.7  & 44$^{+9}_{-8}$	& $6.1\times 10^{-16}$	& $>$28.3	& 23.9$\pm0.4$	& 22.7$\pm0.1$	& 22.4$\pm0.2$	\\
03 32 20.36 & $-$27 42 28.5  & 42$^{+11}_{-10}$	& $8.7\times 10^{-16}$	& $>$28.3& $>$25.9	& $>$25.1	& $>$25.0	\\
03 32 25.83 & $-$27 51 20.3  & 25$^{+7}_{-6}$	& $3.3\times 10^{-16}$	& $>$28.4	& $>$25.9	& $>$25.1	& 23.3$\pm0.4$	\\
03 32 33.14 & $-$27 52 05.9  & 44$^{+9}_{-7}$	& $5.9\times 10^{-16}$	& $>$28.4& 25.7$\pm1.1$	& 24.8$\pm0.5$	& 25.4$\pm1.5$	\\
03 32 51.64 & $-$27 52 12.8  & 45$^{+11}_{-9}$	& $1.3\times 10^{-15}$	& $>$28.4& $>$25.9	& 24.1$\pm0.3$	& 23.5$\pm0.6$
\enddata
\tablenotetext{a}{Positions for the X-ray sources, from
	Alexander et al. (2003).}
\tablenotetext{b}{This source was also optically undetected by
	Yan et al. (2003);
note that we detect their other sources in our $z_{850}$ data.}
\end{deluxetable}

\end{document}